\documentclass[prb,twocolumn,showpacs,superscriptaddress]{revtex4}

\usepackage{graphicx}
\usepackage{amsmath}
\usepackage{amssymb}
\usepackage{epsfig}

\renewcommand{\v}[1]{{\bf #1}}

\def\eqa{\begin{eqnarray}}
\def\eea{\end{eqnarray}}
\newcommand{\eq}{\begin{equation}}
\newcommand{\ee}{\end{equation}}
\newcommand{\nn}{\nonumber\\}

\newcommand{\<}{\langle}
\renewcommand{\>}{\rangle}

\newcommand{\ra}{\rightarrow}

\newcommand{\al}{\alpha}
\newcommand{\bt}{\beta}
\newcommand{\del}{\delta}
\newcommand{\Del}{\Delta}

\newcommand{\ga}{\gamma}
\newcommand{\Ga}{\Gamma}

\newcommand{\La}{\Lambda}

\newcommand{\si}{\sigma}

\usepackage{color}
\newcommand {\B}{\textcolor {blue}}

\begin{document}

\title{Functional renormalization group study of superconductivity in doped Sr$_2$IrO$_4$}

\author{Yang Yang}
\affiliation{National Laboratory of Solid State Microstructures $\&$ School of Physics, Nanjing
University, Nanjing, 210093, China}
\author{Wan-Sheng Wang}
\affiliation{National Laboratory of Solid State Microstructures $\&$ School of Physics, Nanjing
University, Nanjing, 210093, China}

\author{Jin-Guo Liu}
\affiliation{National Laboratory of Solid State Microstructures $\&$ School of Physics, Nanjing
University, Nanjing, 210093, China}

\author{Hua Chen}
\affiliation{Zhejiang Institute of Modern Physics $\&$ Department of
Physics, Zhejiang University, Hangzhou 310027, China}

\author{Jian-Hui Dai}
\affiliation{Department of Physics, Hangzhou Normal University,
Hangzhou 310036, China}

\author{Qiang-Hua Wang}
\affiliation{National Laboratory of Solid State Microstructures $\&$ School of Physics, Nanjing
University, Nanjing, 210093, China}

\begin{abstract}
Using functional renormalization group we investigated possible superconductivity in doped Sr$_2$IrO$_4$. In the electron doped case, a $d^*_{x^2-y^2}$-wave superconducting phase is found in a narrow doping region. The pairing is driven by spin fluctuations within the single conduction band. In contrast, for hole doping an $s^*_{\pm}$-wave phase is established, triggered by spin fluctuations within and across the two conduction bands. In all cases there are comparable singlet and triplet components in the pairing function. The Hund's rule coupling reduces (enhances) superconductivity for electron (hole) doping. Our results imply that hole doping is more promising to achieve a higher transition temperature. Experimental perspectives are discussed.
\end{abstract}

%

\maketitle

\section{Introduction}
\begin{figure}
\includegraphics[width=8.5cm]{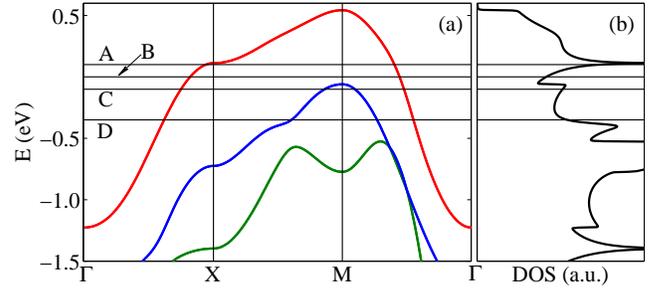}
\caption{ (Color online) (a) The Electronic structure of Sr$_2$IrO$_4$ described by $H_0=H_{Kin}+H_{SOC}$, with $H_{Kin}$ from Ref.\cite{vmc-1,vmc-2}. Each band remains to be doubly degenerated. The horizontal lines indicate Fermi levels addressed in the text. The Fermi energy of the undoped compound is set to zero (line-B). (b) Normal state density of states.}\label{band}
\end{figure}

Recently, the iridium oxide Sr$_2$IrO$_4$ has been subject to extensive investigations.~\cite{Randall,Crawford,Cao,BJKim-1,BJKim-2,SJMoon,xray1,xray2,xray3,Cetin,neutron} In the parent compound the Ir atom is in the $5d^5$ configuration. The spin-orbital coupling (SOC) splits the $t_{2g}$-manifold into filled $J=3/2$ multiplets and half-filled $J=1/2$ doublets, leading to a band structure as shown in Fig.\ref{band}. Since the top $J=1/2$ band is half-filled and the width is narrowed down to the scale of local interactions, the parent compound was argued to be a Mott insulator. Indeed, transport measurements revealed insulating behavior,\cite{Cao} and a canted antiferromagnetic (AFM) order was found in X-ray scattering and neutron diffraction measurements.\cite{BJKim-2,xray1,xray2,xray3,neutron} In analogy to cuprates, an intriguing issue is whether superconductivity (SC) could be realized by doping the parent insulator.\cite{xray2,wangfa}

Theoretically, a variational Monte Carlo (VMC) study of Sr$_2$IrO$_4$\cite{vmc-2} suggests $d$-wave SC may appear but only within a narrow region of electron doping. The absence of SC in the hole doped side is not straightforward to understand. In fact, by sufficient hole doping, both of the two higher bands are cut by the Fermi level (see Fig.\ref{band}), forming Fermi pockets around the $\Ga$ and $M$ points in the Brillouine zone. (In this case the band structure questions the notion of doped Mott insulator for Sr$_2$IrO$_4$.) Instead, the Fermi surface topology is closely similar to that in iron pnictides, where inter-pocket scattering proves to be very efficient to drive $s_\pm$-wave superconductivity.\cite{iron-SC,s+-} However, this does not seem to be the case in the VMC results. Given the unavoidable bias in VMC, we think it beneficial to perform a complementary, yet unbiased search for SC in doped Sr$_2$IrO$_4$.

In this paper we resort to functional renormalization group (FRG).\cite{wetterich} This is because FRG treats all electronic instabilities on equal footing without {\em a priori} assumption of the candidate order parameters. It proves successful in doped cuprates and iron pnictides.\cite{Honerkamp,frg} We limit ourselves to sufficient electron/hole doping so that FRG has a better chance to be reliable, as in the practice for doped cuprates.\cite{CuO} Since the three bands overlap within an energy window of order $1$eV, as seen in Fig.\ref{band}, we include all of the $t_{2g}$ orbitals, and apply the recently developed singular-mode functional renormalization group (SMFRG). \cite{wws1,xyy1,xyy2,wws2,xyy3,yy,sro,wws3}
Compared to the other FRG schemes, it has the additional advantage to deal with orbital and spin degrees of freedom and the SOC among them in a more straightforward manner.

Our main findings are as follows: In the electron doped case, a $d^*_{x^2-y^2}$-wave superconducting phase is found in a narrow doping region close to the van Hove singularity, in agreement to VMC. The pairing is driven by spin-like fluctuations within the single conduction band. In contrast, for hole doping an $s^*_{\pm}$-wave phase is established, triggered by spin fluctuations within and across the two conduction bands. In all cases there are comparable singlet and triplet components in the pairing function. The Hund's rule coupling reduces (enhances) superconductivity in the electron (hole) doped case. In view of reasonable Hund's rule coupling, the doping range and the pairing scale, we propose that hole doping is more promising to achieve a higher transition temperature. Experimental perspectives are discussed.\\

\section{Model and method}
We begin with specification of the model hamiltonian $H$. The free part $H_0$ of $H$ contains the spin-invariant kinetic part, $H_{Kin}$, and an atomic SOC part, $H_{SOC}=-\frac{1}{2}\lambda\sum_j \psi^\dag_j \v L\cdot\v \si\psi_j$, where $\psi_j$ is the annihilation field operator at site $j$, and $\v L$ and $\si/2$ are the operators for the orbital and spin angular momenta. To be specific, the nonzero elements of $\v L=(L_x,L_y,L_z)$ in the orbital basis $(d_{xz},d_{yz},d_{xy})$ are, \eqa L_x^{31}=-L_x^{13}=L_y^{23}=-L_y^{32}=L_z^{12}=-L_z^{21}=i.\eea We take $H_{Kin}$ suggested in Refs.\cite{vmc-2,vmc-1}, where the effect of lattice distortions~\cite{Crawford,wangfa} has been taken into account. For SOC we set $\lambda=0.5$eV. The corresponding band structure and density of states (DOS) for $H_0=H_{Kin}+H_{SOC}$ are shown in Fig.\ref{band}(a) and (b), respectively. (Notice that each band remains two-fold degenerate.) The horizontal line-B corresponds to
the undoped Fermi level, and the other lines to the doped cases to be addressed specifically later.

The interacting part $H_I$ of $H$ contains intra-orbital repulsion $U$, inter-orbital repulsion $U'$, Hund's rule spin exchange $J$ and pair hopping $J'$. The explicit form of $H_I$ is standard and can be found elsewhere.\cite{yy} We apply the Kanamori relations $U = U'+2J$ and $J=J'$ to reduce the number of independent parameters. According to an estimate by constrained random phase approximation,\cite{crpa} we limit ourselves in the parameter ranges $U=2\sim 3$eV and $J/U=0.05\sim 0.20$.

The bare interactions, upon full anti-symmetrization, provide the initial values of the running interaction vertices (versus a decreasing energy scale) in SMFRG. A general interaction vertex function can be decomposed as \eq V^{\al, \bt; \ga , \del}_{\v k, \v k', \v q} \ra \sum_m S_m(\v q) \phi^{\al, \bt}_m(\v k, \v q) [\phi^{\ga, \del}_m(\v k', \v q)]^*, \ee
either in the particle-particle (p-p) or particle-hole (p-h) channel. Here, $(\al, \bt, \ga, \del)$ are dummy labels for orbital and
spin indices, $\v q$ is the collective momentum, and $\v k$ (or $\v k'$) is an internal momentum of the Fermion bilinears $\psi^\dag_{\v k + \v q, \al}
\psi^\dag_{-\v k, \bt}$ and $\psi^\dag_{\v k + \v q, \al} \psi_{\v k ,\bt}$ in the p-p and p-h channels, respectively. The fastest growing eigenvalue
$S(\v Q)$ implies an emerging order associated with a collective wave vector $\v Q$ and eigenfunction (or form factor) $\phi(\v k, \v Q)$.\cite{note}
In the p-p channel $\v Q=0$ is always realized at low energy scale due to the Cooper mechanism. More technical details can be found elsewhere.\cite{xyy1,wws1}\\

\begin{figure}
\includegraphics[width=8.5cm]{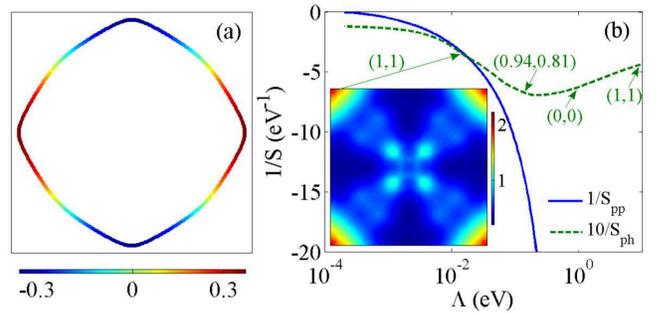}
\caption{(Color online) Results for $n=5.20$. (a) Fermi surface and gap function $\Del(\v k)$ (color scale). (b) FRG flow of $1/S_{ph,pp}$, the inverse of the leading attractive interactions, versus the running energy scale $\La$. Notice that $1/S_{pp,ph}\ra 0^-$ if $S_{pp,ph}$ diverges. The arrows indicate snapshots of the leading momentum $\v Q$ (divided by $\pi$) in the p-h channel. The inset shows $\ln |S_{ph}(\v q)|$ in the Brillouine zone at the final energy scale. }\label{caseA}
\end{figure}

\section{Electron doping} We first discuss the electron doped case with the band filling $n=5.20$, corresponding to line-A in Fig.\ref{band}. The Fermi surface is contributed by the upper band alone, as shown in Fig.\ref{caseA}(a), but we should emphasize that our SMFRG includes virtual excitations from all bands.  Fig.\ref{caseA}(b) shows the FRG flow of the leading eigenvalues $S_{pp,ph}$ versus the running energy scale $\La$ (the infrared cutoff of the Matsubara frequency) for $U=2.4$eV and $J/U=0.055$. Apart from some intermediate deviations the momentum associated with $S_{ph}$ is close to $\v Q=(\pi,\pi)$. The inset shows $S_{ph}(\v q)$ versus $\v q$ at the final energy scale. There is a broad peak around $\v Q$. We checked that the associated form factors describes site-local spins aligned in the plane. Thus AFM spin fluctuations with {\em easy-plane anisotropy} exist. The enhancement of such spin fluctuations can be ascribed to the quasi-nesting of the Fermi surface shown in Fig.\ref{caseA}(a) and the proximity to the van Hove singularity near $X$ (see Fig.\ref{band}). The easy-plane anisotropy is from SOC, and appears to be consistent with the easy-plane AFM order in the parent compound,\cite{BJKim-2,xray1,xray2,xray3,neutron} although FRG can not access the Mott limit.

From Fig.\ref{caseA}(b), As $S_{ph}$ is enhanced below $\La=0.1$eV, it triggers $S_{pp}$ to increase and eventually diverge. Therefore the driving force of pairing here is the AFM spin fluctuation discussed above. We write the (matrix) pairing form factor as \eqa \phi_{pp}(\v k)=(g_{\v k}+\gamma_{\v k})i\sigma_2,\eea with singlet and triplet parts $g_{\v k}$ and $\gamma_{\v k}$, respectively. To describe the momentum dependence, we introduce the lattice harmonics \eqa c_{x}=\cos k_x, \ \ \ c_y=\cos k_y. \label{harmonics} \eea The non-vanishing elements of $g_\v k$ and $\ga_\v k$ in the orbital basis are,
\eqa
&& g_{\v k}^{11/22}\sim ( \mp 0.35\pm 0.20 c_{y/x}\mp 0.08 c_{x/y})\si_0,\nn
&& g_\v k^{33}\sim 0.07(c_y-c_x)\si_0,\\
&& \ga_\v k \sim (0.12 c_x-0.15 c_y)L_x\si_1 + (0.15 c_x-0.12 c_y)L_y\si_2\nn
&& \ \ \ \ \ \ \ + 0.23(c_x-c_y)L_z\si_3. 
\eea
Combining the transformation property of the $d$-orbitals,\cite{wy}  we see $g_\v k$ transforms as $d_{x^2-y^2}$. The symmetry is consistent with the fact that spin fluctuations at the wave vector $\v Q=(\pi,\pi)$ overlap with the $d_{x^2-y^2}$-wave singlet pairing interaction in square lattices. The triplet parts mainly arise from nearest-neighbor bonds, and are orbital-singlets (i.e., odd in orbital space). We notice that $\ga_\v k$ is comparable to $g_\v k$, and is a result of significant SOC. Under point group operations of spin, orbital and momentum, $\ga_\v k$ also transforms as $d_{x^2-y^2}$. According to Ref.\cite{yy} we dub the symmetry of the total pairing function as $d^*_{x^2-y^2}$.  The pairing function respects time-reversal symmetry, which would have been anticipated since the $d$-wave representations on square lattices are non-degenerate. We project the pairing function in the band basis as \eqa \Del_\v k=\<\v k|\phi_{\rm pp}(\v k)(|-\v k\>)^*=\<\v k|g_\v k+\ga_\v k|\v k\>,\eea where $|\v k\>$ is a Bloch state and $|-\v k\>=i\si_2K|\v k\>$ is the time-reversal of $|\v k\>$. The gap function $\Del_\v k$ is shown in Fig.\ref{caseA}(a) (color scale) on one of the doubly degenerated Fermi surfaces, revealing the $d$-wave sign structure consistent with the above symmetry analysis in the spin-orbital basis. We notice that the gap function doesn't change between the degenerate Fermi surfaces. This is because any band dependence is determined by $\<\v k|\ga_\v k|\v k\>$, but $\ga_\v k$ is of the same form of SOC, which nonetheless does not break the degeneracy. We notice in passing that the pairing function in the orbital basis in this paper would be useful in further VMC studies.

\section{Hole doping} 
\begin{figure}
\includegraphics[width=8.5cm]{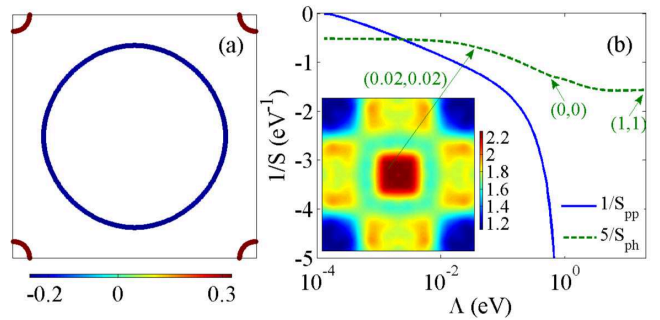}
\caption{(Color online) The same plot as Fig.\ref{caseA}, except that $n=4.83$.}\label{caseB}
\end{figure}

We now discuss the hole doped cases. First consider a band filling $n=4.83$ associated with line-C in Fig.\ref{band}. The Fermi surface topology changes drastically. A large $\Ga$-pocket from the upper band and a small $M$-pocket from the middle band appear, as shown in Fig.\ref{caseB}(a). For reasons to be clearer later, we set $U=2.4$eV and $J/U=0.175$, with a larger Hund's rule coupling. The FRG flow is shown in Fig.\ref{caseB}(b). In this case, the $\v Q$ vector for the leading $S_{\rm ph}$ evolves from $(\pi,\pi)$ at high energy scales to small momenta at moderate and low energy scales. The inset shows $S_{ph}(\v q)$ versus $\v q$ at the final energy scale. Incommensurate peaks around the zone center are obvious. The fact that they are stronger at low energy scales suggests that they arise from intra-pocket scattering around $M$. We checked that such fluctuations are also spin-like, but now the fluctuating spins are aligned along the out-of-plane directions. Thus hole doping leads to ferromagnetic-like spin fluctuations with easy-axis anisotropy. The difference to the electron doped case can be easily checked, e.g., by neutron scattering. On the other hand, there are secondary peaks at $\v Q'\sim (\pi,\pi/4)$ and its symmetry images in $S_{ph}(\v q)$. They are also spin-like by checking the associated form factors. These spin fluctuations can only come from inter-pocket (thus inter-band in our case) scattering. From Fig.\ref{caseB}(b), as spin fluctuations are enhanced in the intermediate energy window, attractive pairing interaction $S_{pp}$ is induced rapidly, and eventually diverge. At this stage, we find the following non-vanishing elements for $\phi_{pp}(\v k)$,
\eqa && g_\v k^{11/22}\sim (0.15 c_{x/y}+0.02 c_{y/x})\si_0, \nn
     && g_\v k^{33}\sim [0.37-0.03(c_{x}+c_{y})]\si_0,\\
     && \ga_\v k\sim 0.07(L_x\si_1+L_y\si_2)\nn
     && \ \ \ \ \ \ +[0.40-0.05(c_{x}+c_{y})]L_z\si_3.
\eea
Symmetry analysis similar to the previous case shows that the gap function transforms as $s$-wave.\cite{wy} The singlet and triplet parts are comparable in magnitude, and both are time-reversal invariant. The projection of $\phi_{pp}(\v k)$, or $\Del(\v k)$, is shown in Fig.\ref{caseB}(a) (color scale). We see that $\Del(\v k)$
is roughly isotropic on each pocket, but changes sign from $\Ga$ to $M$ pocket. Combined with the admixture of singlets and triplets in the orbital basis, we dub the global pairing symmetry as $s^*_\pm$-wave.\cite{spm} For the singlet part the sign change across the pockets enjoys the scattering provided by the secondary spin fluctuations near $\v Q'$ mentioned above. We conclude that pairing is driven by spin fluctuations within the hole-like band, and further enhanced by the inter-pocket scattering in the two conduction bands. The reason that the inter-pocket scattering is not leading is because the electron and hole pockets are poorly nested.

We find the above picture also applies for higher levels of hole doping, except that the wavevector $\v Q$ of the leading spin fluctuations becomes larger (since the hole pocket is enlarged), and $\v Q'$ for the sub-leading ones becomes closer to $(\pi,\pi)$ (since the quasi-nesting between the pockets is improved). Instead of repeating the discussions, we provide the pairing function for $n=4.25$ (in view of potential application in VMC), associated with line-D in Fig.\ref{band},
\eqa && g_\v k^{11/22}\sim [-0.11-0.30(c_x+c_y)]\si_0, \nn
     && g_\v k^{33}\sim [-0.21-0.14(c_{x}+c_{y})]\si_0,\\
     && \ga_\v k\sim (0.17-0.02c_{x}-0.12c_{y})L_x\si_1\nn
     && \ \ \ \ \ \ +(0.17-0.12c_{x}-0.02c_{y})L_y\si_2\nn
     && \ \ \ \ \ \ +[0.19+0.04(c_{x}+c_{y})]L_z\si_3.
\eea obtained under the same parameters $U$ and $J$ as above. The pairing symmetry remains to be $s^*_\pm$-wave. We notice that at this level of hole doping, the hole pocket is quasi-nested, and this leads to stronger intra-pocket spin fluctuations and hence stronger SC (see below).

\begin{figure}
\includegraphics[width=8.5cm]{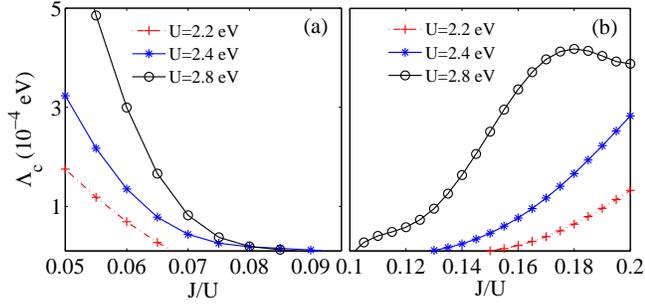}
\caption{(Color online) The superconducting critical scale $\La_c$ versus $J/U$ for various $U$. (a) The $d_{x^2-y^2}^*$-wave pairing at electron doping $n=5.20$. (b) The $s^*_\pm$-wave pairing at hole doping $n=4.83$.}\label{hund}
\end{figure}

\section{Systematics} We have performed systematic calculations by varying the bare interaction parameters. Fig.\ref{hund} shows the critical scale $\La_c$, the energy scale at which the superconducting instability occurs, versus $J/U$ for various values of $U$. For a fixed $J/U$, $\La_c$ increases with $U$. The effect of $J$ for a fixed $U$ is highly nontrivial, however. In the electron doped case, Fig.\ref{hund}(a) shows that the Hund's coupling $J$ suppresses $\La_c$ for $d^*_{x^2-y^2}$-wave pairing in the electron doped case. In the contrary, in the hole doped case $s^*_{\pm}$-wave pairing is enhanced by $J$, as shown in Fig.\ref{hund}(b). The systematics is consistent with the fact that the Hund's rule coupling favors spin fluctuations at smaller wavevectors. Judging from Fig.\ref{hund} we conclude that hole doping is more promising to achieve a higher transition temperature for a reasonable Hund's rule coupling (e.g., $J/U\geq 0.1$.)

\begin{figure}
\includegraphics[width=8.5cm]{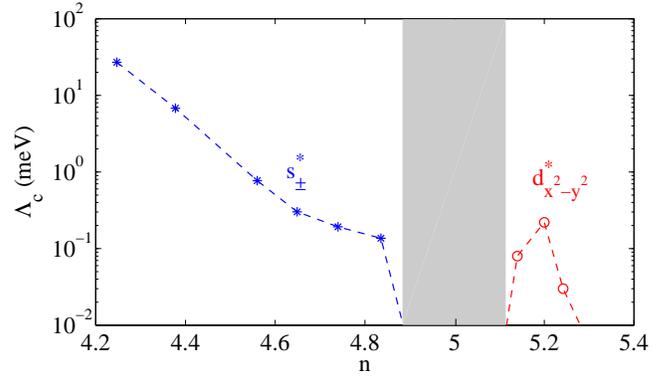}
\caption{(Color online) The superconducting critical scale $\La_c$ versus doping. Here $U=2.4$eV, and $J/U=0.055$ ($0.175$) for electron (hole) doping.}\label{dope}
\end{figure}

On the other hand, we have also performed systematic calculations by varying the filling level $n$. Fig.\ref{dope} shows the $n$ dependence of $\La_c$. The grayed region is not considered since it is too close to the Mott insulating limit for FRG to be reliable. We are interested in sufficient electron/hole doing away from this region. We set $U=2.4$eV here for illustration. In principle we also need to fix $J$ to have a fair comparison between electron and hole doping. However, since $J$ is badly unfavorable in the electron doped case ($n>5$), we set $J/U=0.055$ just in order to have a sizable $\La_c$. Even in this case, SC exists only within a narrow doping region around $n=5.2$ (close to the van Hove filling), in agreement to the VMC result. Instead, in the hole doped case ($n<5$), we set a reasonable value $J/U=0.175$ for definiteness. We see the SC phase extends for all $n\leq 4.83$, and $\La_c$ is enhanced up to $\La_c\sim 30$meV for $n=4.25$. This pairing scale is of the same order of that in iron pnictides, and we conclude that the deeply hole-doped Sr$_2$IrO$_4$ could be a high-$T_c$ superconductor.

\section{Experimental perspectives} 
\begin{figure}
\includegraphics[width=8.5cm]{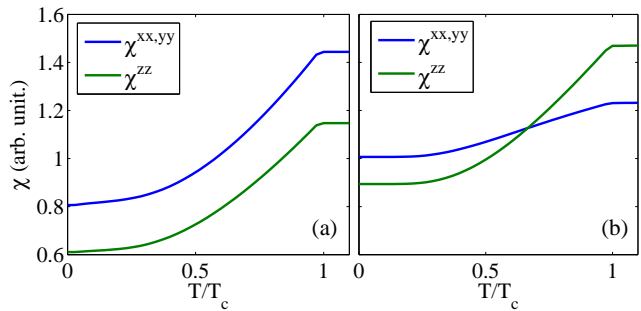}
\caption{(Color online) Spin susceptibilities $\chi^{xx,yy,zz}$ as a function of temperature for (a) $d_{x^2-y^2}^*$-wave pairing in electron doped case $n=5.20$, and (b) $s_\pm^*$-wave pairing in the hole doped case $n=4.83$. Here $T_c$ is the mean field transition temperature. }\label{chi}
\end{figure}

We discuss some experimental consequences regarding the pairing functions obtained so far. Since the $d_{x^2-y^2}^*$-wave pairing has a nodal gap on the Fermi surface, while the $s_\pm^*$-wave pairing is fully gapped, they can be easily differentiated by low temperature thermodynamic measurements (such as the specific heat and superfluid density) and by spectroscopic measurements (such as angle-resolved photoemission and scanning tunneling microscopy). The change of spin anisotropy can be easily probed by neutron diffraction. However, since both types of pairing involve comparable mixing of singlets and triplets, the difference in the spin susceptibility is not as straightforward. We performed mean field calculations in both cases, with the pairing interaction derived from SMFRG (slightly before the divergence scale), and calculated the direction-resolved spin susceptibilities $\chi^{xx,yy,zz}$ versus temperature $T$. The results are shown in Fig.\ref{chi}(a) for $d^*_{x^2-y^2}$- and (b) for $s^*_\pm$-wave pairing for $n=5.20$ and $n=4.83$, respectively. In both cases the susceptibilities are above $40\%$ of the normal state value as $T\ra 0$, and there are anisotropy between $(\chi^{xx},\chi^{yy})$ versus $\chi^{zz}$. Such behaviors, combined with the spectroscopic measurements, would provide an unambiguous probe of the novel pairing functions predicted here.

\section{Conclusions and remarks}
To conclude, in electron (or hole) doped Sr$_2$IrO$_4$, a $d^*_{x^2-y^2}$-wave (or $s^*_\pm$-wave) superconducting phase is possible. They are triggered by in-plane AFM spin fluctuations for electron doping, and by out-of-plane spin fluctuations within the hole pocket as well as from inter-pocket scattering \B{for hole doping}.\cite{s2d} In all cases there are comparable singlet and triplet components. The effect of Hund's rule coupling $J$ suppresses (enhances) SC in the electron (hole) doped region significantly. A reasonable value of $J/U\geq 0.1$ makes hole doping more promising to achieve a higher transition temperature. Experimental perspectives are discussed.

We notice that superconductivity has not been observed yet experimentally by electron doping.\cite{Korneta} While further efforts are needed, {\em our results for hole doping stimulate a new direction.} Experimentally, hole doping can be achieved by substituting K or Na for Sr in Sr$_2$IrO$_4$. Presently isovalent substitution of Ca or Ba for Sr,\cite{Shimura} and partial substitution of Ru for Ir are reported.\cite{Ru4Ir}

\acknowledgments{The project was supported by NSFC (under grant No.11023002 and
No.11274084) and the Ministry of Science and Technology of China (under grant No.2011CBA00108 and 2011CB922101).}


\begin{references}
\bibitem{note} Notice that all $\v Q$-vectors have symmetry images, and could
evolve with the running energy scale.
\bibitem{spm} Apart from the important triplet components, the $s^*_\pm$ pairing is an interesting analogue of $s_\pm$ in iron pnictides.
\end{references}
\end{document}